\documentstyle[12pt,preprint,prl,aps,amsfonts,floats]{revtex}
\tighten
\begin{document}
\title{Soft Photons from Off-shell Particles in a Hot Plasma }
\author{P.A.Henning and E.Quack
       \thanks{e-mail: P.Henning@gsi.de, E.Quack@gsi.de }}
\address{Theory Group, Gesellschaft f\"ur Schwerionenforschung (GSI) \\
        P.O.Box 110552, 64220 Darmstadt, Germany \\
        (May 1995)}
\date{GSI-Preprint 95-29}
\maketitle
\begin{abstract}
Considering the propagation of off-shell particles 
in the framework of thermal field theory, we present the general formalism 
for the calculation of the production rate of soft photons 
and dileptons from a hot plasma. 
This approach is illustrated with an electrodynamic plasma. 
The photon production rate from strongly interacting quarks in the 
quark-gluon plasma, which 
might be formed in ultrarelativistic heavy ion collisions, is calculated in 
the previously unaccessible regime of photon energies of the order of the 
plasma temperature within an effective field theory incorporating 
dynamical chiral symmetry breaking. 
\end{abstract}
The main goal of the present experimental effort invested in ultrarelativistic 
heavy ion collisions (URHIC) is to observe an excursion of the system into 
the phase of a quark-gluon plasma (QGP) \cite{qm}. 
A direct way to `see' this shortlived state would 
be by photons emitted from the hot plasma, see the recent discussion 
\cite{phot}, as well as from dileptons. 
Since these probes interact only electromagnetically, 
this signal is not distorted by later interactions as 
are others. However, it is buried under a 
background of photons from different origin such as $\pi^0$ decays or from 
hadronic reactions \cite{kap1}. 
The knowledge of the thermal spectrum from 
theoretical calculations would thus help to disentangle the various 
sources and to identify the phases reached during the collision. 

This problem represents a challenge to theory in itself, due to the
nonperturbative nature of the photon emission process: Multiple
rescattering of the emitting particles and the Landau-Pomeranchuk-Migdal 
(LPM) effect play an important role in the low energy sector for photon 
energies $E_{\gamma} \leq T$ \cite{wel94}. 

In the present paper, we improve on existing calculations of the photon 
production rate by taking thermal scattering and subsequent off-shell 
propagation of particles into account.
The problem is addressed in the framework of thermal field theory,
results are given for a QED plasma as well as for a QGP within a 
model incorporating dynamical chiral symmetry breaking.

In lowest order, the production of photons proceeds via $q\bar{q}$ 
annihilation and Compton processes. Summing up these two processes with a 
thermal quark ($q$), $\bar{q}$ and gluon distribution of temperature $T$ 
gives the 
production rate $R$ (per unit volume element) to lowest order as  
\begin{equation}
\label{eqralo}
  R^0 = 
  E\frac{dN^0_{\gamma}}{d^3\bbox{p}} = \frac{5}{9} 
  \,\frac{\alpha \alpha_s}{2 \pi^2}\, T^2\, {\mathrm e}^{-E/T}\,
   \left[ \log{\frac{E T}{m^2}} + c^0 \right]
\end{equation}
with some constant $c^0$. This rate diverges when $m \to 0$, 
which is the crucial limit of chiral symmetry restoration for 
strongly interacting quarks approaching the phase transition temperature. 
A shielding of this unphysical divergence requires to calculate
{\em medium effects\/} on the emission process.

The application of the Braaten--Pi\-sarski method of hard thermal loops 
\cite{brpi} to this problem has been studied recently \cite{kap1,bai1}. 
In the rates obtained with this approach, the term in the square 
brackets of eq.(\ref{eqralo}) is replaced by $\log(E/T)$ which clearly 
shows the limitations of the approach to the region of $E_{\gamma} \gg T$.

The dominant physical process for quarks emitting soft photons with energies 
$E_{\gamma} \leq T$ is the scattering in the medium, 
which  results in an energy uncertainty 
as the quark propagates. Formally, we describe this off-shellness by a 
finite width of the quark in analogy to the decay width of an 
excited state. Now, after an interaction with the medium which 
sets the quark off-shell, instead of interacting again with the medium the 
quark may also emit a real photon. Thus, 
the quark width is directly related to the emission rate 
of soft photons. Taking into account such a spectral width 
naturally removes the infrared divergences mentioned before, 
and enables us to give production rates for soft photons.

We now present a brief summary of the formalism. 
As argued above, effective 
field theory in a hot system requires the use of physical states with a 
nonzero width \cite{L88,h94rep}.
It is calculated from the imaginary part of the retarded quark self energy 
$\Sigma^R_q(p)$. The real part of this self energy function is absorbed
into the mass of the quark, by approximation of 
\begin{equation}
\label{lop}
\left(S^R(p)\right)^{-1} = p_\mu\gamma^\mu - m_q^0 - \Sigma^R_q(p) \approx 
 (p_0\pm {\mathrm i}\gamma_q)\gamma_0 - \bbox{p}\bbox{\gamma} - m_q
\end{equation}
in the vicinity of $p_0=\pm\sqrt{\bbox{p}^2 + m^2_q}$. This equation also
determines the quark width $\gamma_q$, which then enters  
the spectral function ${\cal A}(E,\bbox{p})$. This function, which is
up to a factor the imaginary part of the retarded
full propagator of the quark field, 
deviates from its vacuum (perturbative) form 
$( E\gamma^0 + \bbox{p}\bbox{\gamma} + m_q)\,
 \mbox{sign}(E)\,\delta(E^2-\bbox{p}^2 - m^2_q)$. With the above approximation 
we obtain
\begin{eqnarray}
\label{eqaq} 
{\cal A}_q(E,\bbox{p}) &=& \frac{\gamma_q}{\pi} 
\frac{\gamma_0\left(E^2 + \Omega_q(\bbox{p})^2\right) -
      2 E \bbox{\gamma}\bbox{p} + 2 E m_q}{
  \left(E^2 - \Omega_q(\bbox{p})^2\right)^2 + 4 E^2 \gamma_q^2}
\;.\end{eqnarray}
Here, $\gamma_\mu = (\gamma_0,\bbox{\gamma})$ is the four-vector of
Dirac matrices and
 $\Omega_q(\bbox{p})^2 = \bbox{p}^2 + m^2_q + \gamma_q^2$.
One may regard this spectral function as the generalization of the
standard energy-momentum relation $E^2 = \bbox{p}^2 + m^2_q$ to a broader 
distribution for thermally scattered particles, in this particular 
case represented by a double Lorentzian. Retarded as well as causal 
propagators are defined as dispersion integrals of ${\cal A}_q$, and 
we refer to \cite{h94rep} for a formulation of thermal field
theory in terms of spectral functions. 

In a loop expansion, the one-loop (Fock) diagram is the lowest order term 
with a nonvanishing imaginary part. In the following, we restrict ourselves 
to this lowest order. We consider a model where quarks are coupled to 
different types of bosons, to be specified later. The calculation of 
the Fock self energy with full propagators is straightforward \cite{h94rep} 
and gives 
\begin{eqnarray}
\label{eqimsi}
&&{\mathrm Im} \Sigma^R(p_0,\bbox{p})  =\\
\nonumber
&&\;\;\;\;-\pi\,\int\!\!\frac{d^3\bbox{k}}{(2\pi)^3}\,
   \int\limits_{-\infty}^\infty\!\!dE\;\Gamma_{\mu}\,
    {\cal A}_q(E,\bbox{k})\,\Gamma_{\nu} \;
{\cal A}^{\mu\nu}_B(E-p_0,\bbox{k}-\bbox{p})\,\;
    \left(n_q(E)+n_B(E-p_0)\right)
\;.\end{eqnarray}
Here, ${\cal A}_B$ is the boson spectral function, $\Gamma_\mu$ and 
$\Gamma_\nu$ are the interaction matrices at the vertices, 
and $n_B$ ($n_q$) is the standard thermal equilibrium 
Bose (Fermi) distribution functions at temperature $T$.

The width calculated from the quark self energy diagram now enters the photon 
polarization $\Pi$ at finite temperature. The imaginary 
part of the retarded one-loop polarization function $\Pi^R$ is \cite{h94rep}
\begin{eqnarray}
\nonumber
{\mathrm Im} \Pi^R_{\mu\nu}(k_0,\bbox{k}) &  = &
 -\pi\, e^2_q\,\int\!\!\frac{d^3\bbox{p}}{(2\pi)^3}\,
   \int\limits_{-\infty}^\infty\!\!dE \\ 
&&     \mbox{Tr}\left[\gamma_{\mu} {\cal A}_q(E+k_0,\bbox{p}+\bbox{k})
    \gamma_{\nu} {\cal A}_q(E,\bbox{p}) \right]\,\left(n_q(E)-n_q(E+k_0)\right)
\;,\label{eqimpi}
\end{eqnarray}
where $e_q$ is the electric charge of the quark. 
However, this one-loop polarization tensor violates current conservation
and gauge invariance: $k_{\mu}\Pi^{\mu \nu} \neq 0$ and therefore the
the standard sum over the photon polarizations
$\epsilon_{\mu} \epsilon_{\nu} \Pi^{\mu \nu} = \Pi_{\mu}^{\mu}$ 
does not give a meaningful photon production rate, nor is it gauge
invariant.

This can be traced back to the fact, that for an effective field theory
the conserved electromagnetic current is different from the naive
$\overline{\psi} \gamma_\mu\psi$. Due to the the nonlocal nature of the 
effective lagrangian of such a model, the conserved current 
associated with local gauge invariance
acquires a correction term, which makes the current-current correlation
function $\widetilde{\Pi}^{\mu \nu}$ different from the above polarization
tensor, such that $k_{\mu}\widetilde{\Pi}^{\mu \nu} = 0$. 
The correction is of higher loop order than $\Pi^{\mu\nu}$ and 
amounts to the introduction of a vertex correction into eq.
(\ref{eqimpi}).

However, in our approach
we consider the particular case of a quark width which is only dependent on 
the temperature $T$ but not on the quark momentum, as appropriate for 
a slow quark embedded in the medium. In this case, only the $j^0$ component of 
the current is modified and correspondingly only the components 
$\widetilde{\Pi}^{0\nu}=\widetilde{\Pi}^{\nu 0}$ are different from the
one-loop result. It is crucial to realize that the space-like components 
are not modified, $\widetilde{\Pi}^{ij} = \Pi^{ij}$. 
With constant retarded quark self energy, and photon momentum
$k_\mu = (k,0,0,k)$ the gauge invariant 
sum over the polarizations thus reduces to 
$
 \widetilde{\Pi}_{\mu}^{\mu} = \widetilde{\Pi}^{00} - \widetilde{\Pi}^{ii}
 = \widetilde{\Pi}^{00} - \Pi^{ii} = -(\Pi^{11}+\Pi^{22})
$.
In particular, this may be calculated with the unmodified polarization tensor 
from eq.(\ref{eqimpi}).  The photon emission rate out of the hot 
plasma then is
\begin{equation}
 R(E_{\gamma},T) = E_{\gamma} \frac{ dN_{\gamma}}{d^3\bbox{p}} = 
 2\frac{n_B(E_\gamma,T)}{8 \pi^3}\,
 \mbox{Im}\left(\Pi^R_{11}+\Pi^R_{22}\right)
=\frac{{\mathrm i}}{8 \pi^3}\,\left(\Pi^<_{11} + \Pi^<_{22}\right)
\;.\label{prate}
\end{equation}
In the present work we focus on soft photons emitted from a QGP. 
Dynamical chiral symmetry 
breaking and its restoration at the phase transition temperature $T_c$ 
plays an important role and has to be incorporated in a realistic 
description of the quark dynamics. We do so by considering the 
Nambu--Jona-Lasinio model \cite{njl} in the SU(2) version on the quark 
level, see \cite{san1} for a review and the notations used in the following.
  
This effective field theory
models the chiral symmetry properties of QCD in the nonperturbative 
regime by a quartic self-interaction of quarks. At small temperature,
the dominant contribution to the quark self energy is the 
tadpole (Hartree) term, which is expressed in terms of the spectral function
as
\begin{equation}
 \Sigma^H = - 2 G N_C N_f\, \int\!\!\frac{d^3\bbox{p}}{(2\pi)^3}\,
  \int\limits_{-\infty}^\infty\!\!dE\, 
  \mbox{Tr}\left[{\cal A}_q(E,\bbox{p})\right] \, n_q(E) \; .
 \label{eqmh}
\end{equation}
Like any nonrenormalizable model, the NJL requires a momentum cutoff 
$\Lambda$, which can be motivated as a crude incorporation of asymptotic 
freedom at large $Q^2$. For the present generalization, this cutoff
is shifted to the energy integration,
$\Lambda_E = \sqrt{\Lambda^2 + m^2_q(T)}$ for the above diagram,
see \cite{qh95njl} for details.

Usually, the temperature dependent quark mass $m_q(T)$ is the solution 
of the gap equation $m_q = m_0 + \Sigma^H(m_q)$. 
With appropriate parameters, this describes the 
scenario of spontaneous chiral symmetry breaking, i.e., 
the transition from a current quark mass $m_0 \approx 5$ MeV 
to the constituent quark mass $m_q \approx 1/3\; \times$ 
the nucleon mass, and its restoration at a transition temperature $T_c$. 
The only parameters were chosen as
$m_0 = 5$ MeV, $\Lambda = 0.65$ GeV and $G = 4.73\; \mbox{GeV}^{-2}$, 
and result in $T_c = 160$ MeV and $m_{\pi} = 137$ MeV. 

The Fock self energy is the next-to-leading 
order contribution in a $1/N_c$ expansion \cite{ncp}. 
Since we consider quarks with three-momentum $\bbox{p}=0$, 
we can decompose the Fock contribution to the self 
energy in a (complex) scalar and a vector part as 
$\Sigma^{\mbox{\footnotesize Fock}} = \Sigma^{S} + \gamma_0 \Sigma^{V}$. 
These are added to the Hartree self energy, and instead
of the gap equation we solve eq. (\ref{lop}) for the mass and width
of the effective quark field.
\begin{figure}[t]
\vspace*{85mm}
\includegraphics{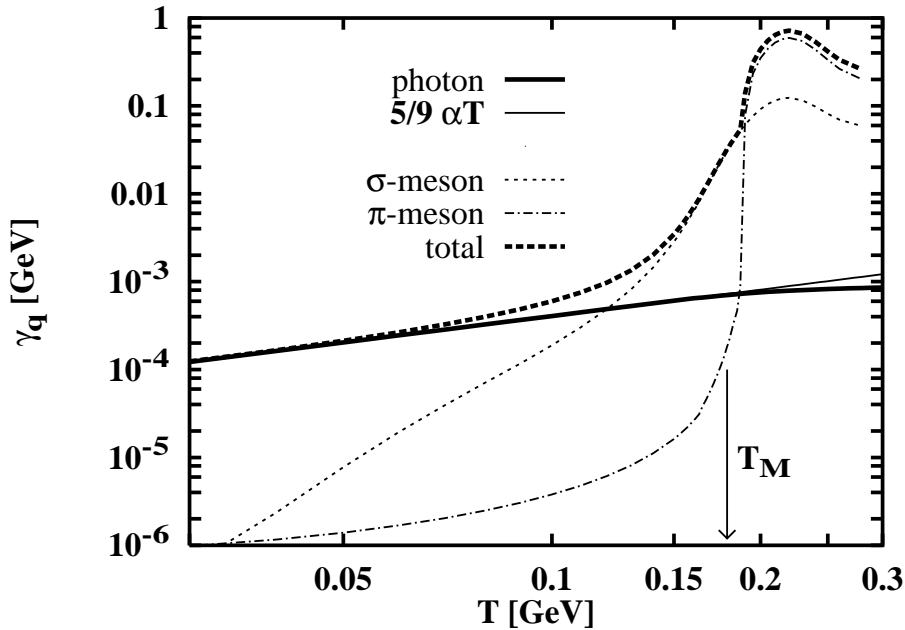}
\caption{Contributions to the width $\gamma_q$ of a quark embedded in a
 QED and QCD plasma as a function of the temperature $T$.}
\end{figure} 

As an illustrative example, we first consider a coupling to
photons only, with a $T$-dependent mass from the NJL model. We perform
this calculation self-consistently using real and imaginary part 
of the Fock self energy (\ref{eqimsi}) for massless free 
photons. This amounts to the multiple scattering and  
LPM effect mentioned before within a QED plasma.

The quark spectral width as function of 
temperature is plotted in fig.~1. Due to the smallness of the coupling
constant one may approximate it very well by the lowest order
which gives 
\begin{equation}
 \gamma_q^{em}(T) \approx \frac{5}{9} \alpha T \cdot 
 [1 - \frac{{\mathrm Re}\Sigma^{V}}{m_q}] \approx \frac{5}{9} \alpha T 
 \cdot 
 [1 - \frac{10}{9} \frac{\alpha}{\pi} \frac{\Lambda T}{m_q^2} ]
\sim \frac{5}{9}\,\alpha \, T
\;.\end{equation}
The photon production rate we obtain from eq. (\ref{prate}) 
with this width is shown in fig.~2 for various typical values of the 
temperature. For small photon energies, i.e.~very soft photons, we find a 
saturation of the rate below values of $E_{\gamma} = 2 \gamma_q^{em}$. 

The physical interpretation of this effect is obvious: The emission of
low-energy photons requires unperturbed propagation of the emitter
over the wavelength of the photon. Along its path however the quark is 
subject to thermal perturbations -- and this hinders the photon emission 
for for $E_\gamma < 2 \gamma_q$. Our result agrees with the conjecture of 
Weldon \cite{wel94}, and we could clarify the dominant suppression scale to be 
twice the spectral width of the emitting particle.

The photon production rate may be approximated as
\begin{equation}\label{cuo}
  R^{\gamma}_{\mbox{\tiny fit}} = \frac{ 4 \gamma_q }{E^2_{\gamma} + 
  4 \gamma^2_q}\, \mbox{e}^{-E_{\gamma}/T}\,z[T]\;\;\; , \;\;z[T]
  \propto \left\{ {\array{lll} T^2 & \mbox{for} & E_{\gamma} \ll 2 \gamma_q\\
                               T   & \mbox{for} & E_{\gamma} \gg 2 \gamma_q
                   \endarray} \right.
\;.\end{equation}
For all temperatures, the limit $E_{\gamma} \to \infty$ is determined 
by the Boltzmann factor $e^{-E_{\gamma}/T}$. 
\begin{figure}[t]
\vspace*{85mm}
\includegraphics{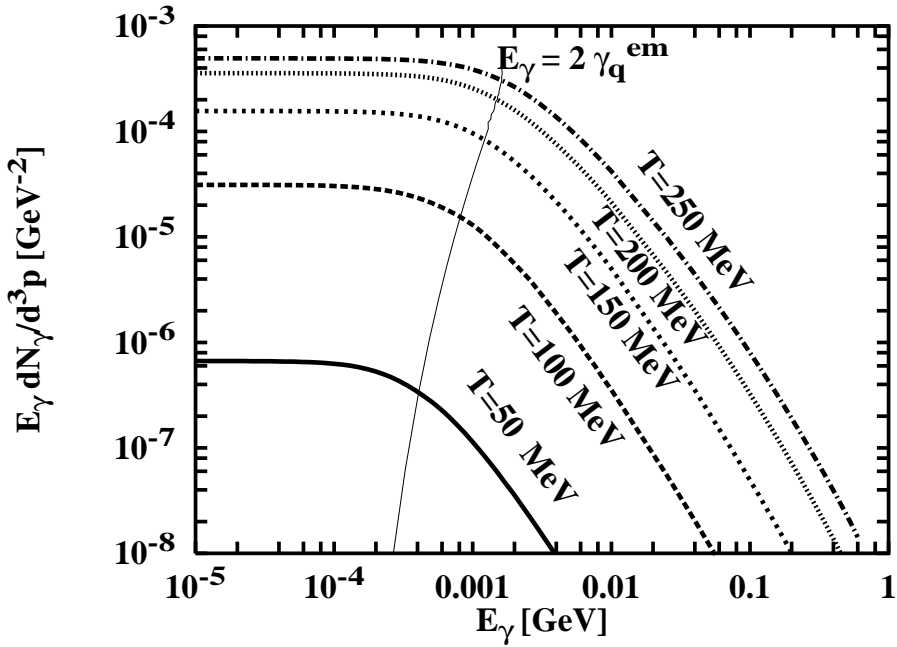}
\caption{Photon production rate $R_{\gamma}$ from a QED plasma as a 
function of the photon energy $E_{\gamma}$.}
\end{figure} 

Finally, we calculate the soft photon production rate from a QGP.
Now, the boson in the Fock self energy diagram of the quark is a collective 
strongly interacting excitation of the system. Within the NJL model, these 
are regarded as effective, $T$-dependent $\pi$ and 
$\sigma$ mesons. In addition to the standard NJL results, we consider also 
the temperature dependent finite width of these effective 
mesons and describe them by generalized Lorentzian spectral functions. 
Both mass and width of the respective meson are determined by the complex 
dispersion relation, as in the fermionic case, eq.(\ref{lop}).  

We do not treat the mesons self-consistently: Only their contribution 
to $\gamma_q$ is considered, their influence on the quark mass
is neglected \cite{ncp}. Physically, our approach amounts to consider
photon emission processes, which are initiated by the interaction of the
quark with a {\em single\/} hot meson. The resulting quark width $\gamma_q$ 
is plotted in fig.~1. For low temperatures, we again find 
$\gamma \propto T$ for each of the mesonic channels. Due to the 
quasi--Goldstone mode of the pion, its contribution remains 
negligible up to the Mott temperature $T_M$ = 179 MeV, which is defined by 
$m_{\pi}(T_M) = 2 m_q(T_M)$ as the point where the pion can 
dissociate in a $q\bar{q}$ pair. For $T > T_M$, the pionic contribution to 
the quark spectral width is actually dominant. The strong increase in 
$\gamma_q$ around $T_M$ is a result of critical opalescence of the system in 
this temperature range. Towards higher temperatures, the competing effect of 
an increase of the mass of the $\pi$ (now a resonance) again turns the width 
down. 

The results for the photon emission rate, eq.(\ref{prate}), are similar to 
those of fig.~2, apart from the much higher saturation scale 
$\gamma_q\gg \gamma_q^{em}$ at temperatures $T>T_M$.
In fig.~3, we show the photon emission rate at three different
photon energies as a function of temperature. 
Comparing the electromagnetic case to the model including the
quark-meson interaction, we find a surprising result: In the region of the
chiral phase transition, the low-energy photon production rate {\em drops\/} 
with increasing temperature. Eventually the radiation rate is
degenerate for all energies $E_\gamma<2\gamma_q$ (see the flat behavior
of the curves in fig. 2).  In view of eq.~(\ref{cuo}), this is 
understood as a dominance of the saturation effect over the increase 
of temperature. 
\begin{figure}[t]
\vspace*{85mm}
\includegraphics{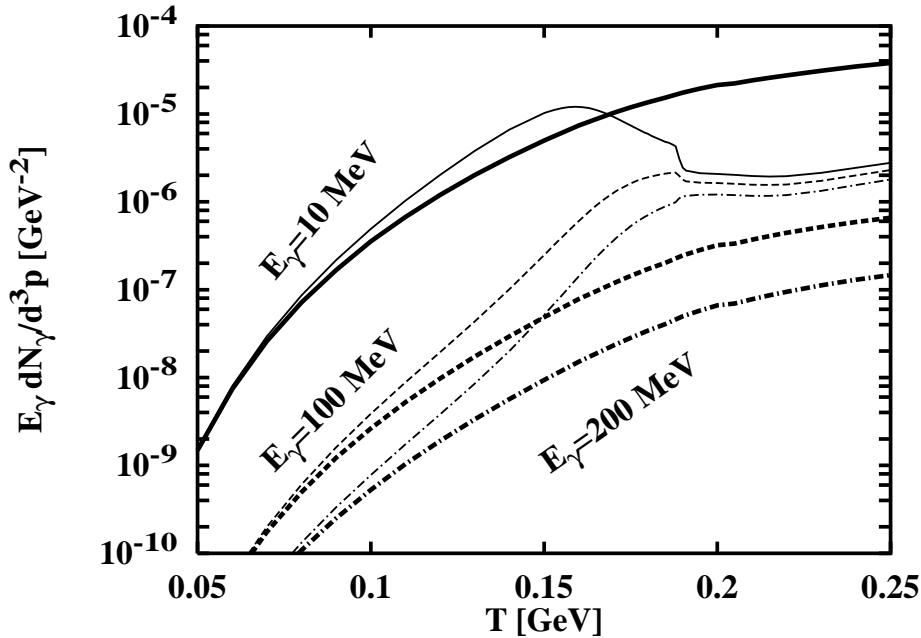} 
\caption{Photon production rate $R$ from a QGP as function of
temperature $T$. Thick lines: $\gamma_q$ purely electromagnetic,
thin lines: meson contributions from fig. 1 added.}
\end{figure} 

Let us emphasize that the qualitative properties of the soft photon rates, 
such as the saturation effect towards low temperatures, follow from general 
physical considerations as we discussed and are in particular independent of 
the particular model we used. The decrease of the production rate of soft 
photons with temperature in the 
region of the phase transition might have important observable consequences. 

Finally, we remark that the results presented for photons are immediately 
related to the rates for the production of soft dileptons from a hot plasma 
by use of the soft photon approximation \cite{ruck}, 
$E^+ E^- dN^{\gamma \gamma \to l^+ l^-}/d^3\bbox{p}^+ d^3\bbox{p}^- = 
\alpha/(2\pi^2 M^2) E dN^{\gamma}/d^3\bbox{p}$. Thus, the present work can be 
applied to a calculation of soft dilepton yields from URHIC, which is of great 
experimental relevance \cite{ceres}. 

\vspace*{.5cm}

{\em Acknowledgements:}\\
Throughout of the course of this work, we profited from discussions with 
J.H\"ufner, J.Knoll and B. Friman.


\end{document}